\documentclass[twocolumn,preprintnumbers,amsmath,amssymb]{revtex4}
\usepackage{graphicx} 
\usepackage{setspace} 
\usepackage{dcolumn} 
\usepackage{bm} 
\usepackage{natbib}
\usepackage{hyperref}
\hypersetup{colorlinks=true,linkcolor=blue,citecolor=blue,urlcolor=blue}
\usepackage{epsfig}
\usepackage{amsmath}
\begin{document}
\title{Structure–Property Correlations in Sb, Ge, and Ga Doped AlFe$_2$B$_2$ for Magnetocaloric Applications}
\author{R. Preyadarshini$^a$}
\author{S. Kavita$^a$}
\author{Ashutosh Kumar$^b$}
\author{D. Sivaprahasam$^a$\footnote{Email: sprakash@arci.res.ac.in}}

\affiliation{$^a$Centre for Automotive Energy Materials (CAEM), International Advanced Research Centre
for Powder Metallurgy and New Materials (ARCI), IITM Research Park, Taramani, Chennai – 600 113, India}
\affiliation{$^b$Functional Materials Laboratory, Department of Materials Science and Metallurgical Engineering, Indian Institute of Technology Bhilai - 491 002, India}
\date{\today}
\begin{abstract}
This study investigates the effects of Sb, Ge, and Ga doping in AlFe$_2$B$_2$ on magnetic and magneto-caloric properties. Samples of AlFe$_2$B$_2$ and AlFe$_{1.9}$M$_{0.1}$B$_2$ (M= Ge, Ga and Sb) with 20\% excess Al were synthesized by arc melting, and the powder processed were investigated for their phase constituents, microstructure, magnetic and magneto-caloric effect. The parent compounds prepared showed the AlFe$_2$B$_2$ phase with a FeB secondary phase. However, in Sb and Ga-doped samples, an additional impurity phase, Al$_{13}$Fe$_4$, was observed apart from FeB, while in Ge-doped, only the AlB$_2$ impurity phase was present. The Curie temperature of AlFe$_2$B$_2$ is 277 K, increasing with Sb, Ge, and Ga doping to 287\,K, 297\,K, and 296\,K, respectively. The magnetization ($M$) is also higher with Ge and Ga addition in the 100-300\,K range; however, with Sb doping, the $M$ decreases significantly compared to parent AlFe$_2$B$_2$. The magnetic entropy change under 2\,T reached 2.93 JKg$^{-1}$K$^{-1}$ near 274\,K in AlFe$_2$B$_2$, which decreases to 2.53 JKg$^{-1}$K$^{-1}$ and 1.92 JKg$^{-1}$K$^{-1}$ with Ge and Ga, respectively. With Sb doping, the MC change was affected dramatically to 0.32 JKg$^{-1}$K$^{-1}$. However, the relative cooling power of Ge doped is the same as that of parent AlFe$_2$B$_2$. This research advances the understanding of the relationship between doping elements and magnetic properties in AlFe$_2$B$_2$ and opens pathways for designing magneto-caloric materials with tailored magnetic characteristics.\\                
\end{abstract}
\maketitle
\section{Introduction}
The magnetocaloric (MC) effect is the foundation for promising sustainable technologies for cooling, such as magnetic refrigeration. It offers an eco-friendly substitute for traditional vapor-compression cooling methods and is estimated to achieve good energy efficiencies.\cite{R1}. The other advantage is that it eliminates harmful synthetic refrigerants like chlorofluorocarbons (CFCs), hydrofluorocarbons (HFCs), and hydrochlorofluorocarbons (HCFCs). This transition can align with the increasing need for cleaner and more efficient cooling solutions.\cite{R2} The application and removal of the magnetic field induce adiabatic temperature change, which can be used effectively in refrigeration. The important characteristics of MC material are large magnetic entropy changes ($\Delta$Sm) over a broad temperature range. Based on the magnetic transition, the MC alloys are classified as first-order magnetic transition material (FOMTM) and second-order magnetic transition material (SOMTM). MC materials with large $\Delta$S$_m$ over a wide range of operational temperatures and high RCP are ideal for efficient cooling applications. 
Earlier, research on MC materials focused on FOMTM such as Gd$_5$Ge$_2$Si$_2$, MnFeP$_{0.45}$As$_{0.55}$, La(Fe$_{1-x}$Si$_x$)13 and Y$_{0.4}$Gd$_{0.6}$Co$_2$ that shows good $\Delta$S$_m$ near room temperature.\cite{R3,R4,R5,R6} These FOMTM compounds have large thermal and magnetic hysteresis, which restricts the applications as it would reduce the efficiency of magnetic refrigeration.\cite{R7} Composition or microstructure optimization can greatly reduce the hysteresis and exhibit good MCE. Since these compounds contain expensive rare-earth elements, using them for large-scale commercial refrigeration systems has its own limitations. Hence, it is necessary to develop earth-abundant and cost-effective MC materials. Due to the limitations associated with the first-order phase transition, there has been an increased focus on magnetocaloric materials that exhibit second-order phase transitions. In this context, AlFe$_2$B$_2$ is a potential material that shows good MC properties around room temperature with reported $\Delta$S$_m$ values (8 JKg$^{-1}$K$^{-1}$ under 2T) that can be a good alternative to existing MC materials.\cite{R8,R9, R10} 
AlFe$_2$B$_2$ is an orthorhombic intermetallic compound with a \textit{Cmmm} space group. Its layered structure features Fe$_2$B$_2$ slabs separated by an Al layer, suppressing direct Fe-Fe and Fe-B interactions. This compound undergoes a ferro-paramagnetic transition near room temperature, with the Curie temperature ranging from 274\,K to 320\,K. The compound exhibits $\Delta$S$_m$ values of 2.9-3.4 JKg$^{-1}$K$^{-1}$ with 2\,T field, attracting considerable research interest to further improve by various methodologies like doping/substituting Al and Fe in the crystal.\cite{R10, R11,R12,R13,R14,R15} Barua et al. studies on MCE of AlFe$_2$B$_2$ with 10\% excess Al addition state that by doping of Ga and Ge, the formation of secondary phase FeB is avoided and reports an increased $\Delta$S$_m$.\cite{R12} In AlFe$_2$B$_2$ synthesis by melt route, obtaining pure phase is highly challenging. Undesirable secondary phases FeB and Al$_{13}$Fe$_4$ that affect its MC characteristics inherently form due to the peritectic reaction occurring during solidification [xx]. Hence, close control of composition and process conditions is critical to minimize the impurity phase. Alternative processes like reactive hot pressing and hot isostatic pressing are being explored to avoid these impurity phases.\cite{R16,R17} Recently, Da Igreja et al. studied the magnetocaloric property of Al$_{1.2}$Fe$_2$B$_2$ synthesized by reactive isostatic pressing where they achieved $\Delta$S$_m$=2.9 JKg$^{-1}$K$^{-1}$ for applied field change $\Delta$H=2\,T with RCP value of 39 J.kg$^{-1}$ at 2\,T.\cite{R17}
This work reports the magnetic and magnetocaloric properties of AlFe$_2$B$_2$ doped with Sb, Ge and Ga prepared by melt route. The alloy samples were prepared by arc melting with 20\% excess Al-containing feedstock over Al:Fe: B 1:2:2 and processed further by powder consolidation. The constituent phases of the prepared alloys, microstructure, $\Delta$S$_m$ and relative cooling power (RCP) were investigated in detail, and the effect of doping is discussed.\\
\section{Materials and Methodology}
\subsection{Materials Synthesis}
\begin{figure*}
\centering
  \includegraphics[width=0.7\linewidth]{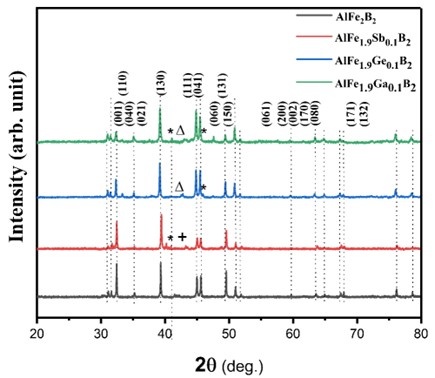}
  \caption{XRD Pattern of Al$_{1.2}$Fe$_2$B$_2$ and Al$_{1.2}$Fe$_{2-x}$M$_x$B$_2$ (M=Ag, Ni, Sb, Ge, and Ga), $+$ - Al$_{13}$Fe$_4$, $*$-FeB, $\Delta$-AlB$_2$.}
  \label{fig1a}
\end{figure*}
\begin{figure*}
\centering
  \includegraphics[width=0.8\linewidth]{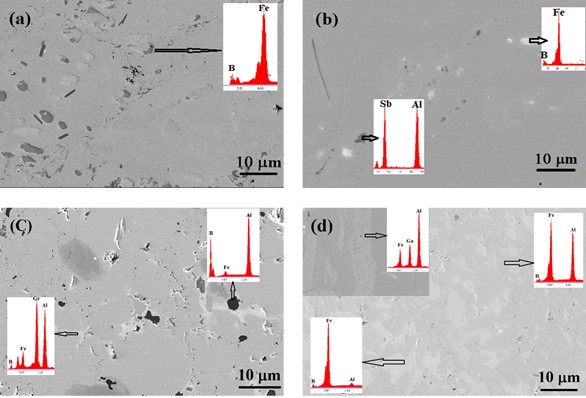}
  \caption{SEM micrographs of (a) Al$_{1.2}$Fe$_2$B$_2$ (b) Al$_{1.2}$Fe$_{1.9}$Sb$_{0.1}$B$_2$ (c) Al$_{1.2}$Fe$_{1.9}$Ga$_{0.1}$B$_2$ and (d) Al$_{1.2}$Fe$_{1.9}$Ge$_{0.1}$B$_2$}.
  \label{fig2a}
\end{figure*}
The Al$_{1.2}$Fe$_2$B$_2$, Al$_{1.2}$Fe$_{1.9}$M$_{0.1}$B$_2$(M=Sb, Ge, Ga) compounds were synthesized by vacuum arc melting (VST, Israel) of Al (99.9\%, American elements, USA), Fe$_2$B (99.9\%, American elements, USA), B (99.5\%, 20 mm, Alfa Aesar, USA), Sb (99.99\%, Thermo Fischer Scientific, India), Ga (99.9\%, American elements, USA) and Ge (99.9\%, Thermo Fisher Scientific, India) weighed according to the stoichiometric composition. Before melting, the chamber was evacuated to below 5$\times$10$^{-5}$ mbar vacuum and flushed with a continuous 99.999\% pure argon flow at 300\,mbar.
Each sample weighing 10\,g was arc melted three times after breaking the vacuum and flipping each melting. The solidified samples were annealed for 96\,h at 1323\,K after sealing under 99.999\% argon in a quartz ampoule, crushed and ground to fine powder in agate pestle and mortar. Using a graphite die, the powder was consolidated in a vacuum hot press (Thermal Technology, USA) at 1273\,K for 2\,h under 50\,MPa pressure to 20\,mm diameter pellets. During hot pressing the chamber vacuum was less than 8$\times$10$^{-5}$mbar. The hot-pressed pellets were cut and polished to study various materials and magnetic characteristics.\\
\subsection{Characterization Details}
The structural characterization of the synthesized powders was performed using a Rigaku Smart Lab powder X-ray diffractometer (Cu-K$\alpha$, $\lambda$=0.15406 nm) with a step of 0.01$^\circ$ and a scan range from 20–80$^\circ$ at 1$^\circ$ per minute. The X’pert high score plus software was used for profile fitting. The lattice parameters ‘\textit{a}’, ‘\textit{b}’ and ‘\textit{c}’ of the orthorhombic AlFe$_2$B$_2$ phase were calculated from d(200), d(060), and d(001) diffraction peaks. The microstructural observation of the hot-pressed pellets was done using a field emission scanning electron microscope (FE–SEM; Zeiss Merlin, Germany). The composition of constituent phases in the samples was analyzed using energy-dispersive X-ray spectroscopy (EDS).  The magnetization as a function of temperature M(T) and applied field M(H) were measured using a physical properties measurement system (PPMS-Quantum Design, Dynacool-9T). The M vs. T measurements were performed from 100\,K to 330\,K at the constant field of 0.05\,T with a temperature sweep rate of 12 K/min. Magnetic isotherms M(H) were recorded within the range of 250\,K-330\,K with a step size of 3\,K from 1 to 7\,T. The M(T) measurements were repeated two times to assess the reliability of the results.\\
\section{Results and Discussion}
\subsection{Phases and Microstructure}
AlFe$_2$B$_2$ crystallizes in the orthorhombic \textit{Cmmm}-type structure. Fig.~\ref{fig1a} shows the XRD results of Al$_{1.2}$Fe$_2$B$_2$ and Al$_{1.2}$Fe$_{1.9}$M$_{0.1}$B$_2$ (M=Sb, Ge and Ga) samples. The diffraction pattern of all samples shows peaks corresponding to the AlFe$_2$B$_2$ primary phase. The constituent phase fraction estimated by Rietveld analysis shows AlFe$_2$B$_2$ (98-000-7593) with 6.9\% of the FeB (98-001-4125) phase. Reported results indicate that increasing the excess Al can further reduce the FeB impurity. However, this could also lead to Al$_{13}$Fe$_4$ in AlFe$_2$B$_2$ and other secondary phases in the doped AlFe$_2$B$_2$.\cite{R18} Hence, optimizing excess Al addition is critical for minimizing the secondary phase content, and between 20-40\% is ideal.\cite{R7,R8} In this work, 20\% excess Al is used, considering that a higher percentage favors multiple impurity phases. In AlFe$_{1.9}$Sb$_{0.1}$B$_2$, the FeB was reduced to 0.58\%; however, an additional small peak (marked as $+$ in Fig.~\ref{fig1a}) corresponding to Al$_{13}$Fe$_4$ was observed. In addition, unknown peaks appear at 40.6$^\circ$ and 40.9$^\circ$, which cannot be assigned to any known phases of the primary material or its common impurities. The Ga-doped sample shows a significant fraction of the FeB phase. The Al$_{1.2}$Fe$_{1.9}$Ge$_{0.1}$B$_2$ shows no diffraction peak corresponding to FeB, and only the AlB$_2$ impurity phase peak is present. The lattice parameters calculated from the XRD peaks of all these compounds are given in Table~\ref{Table I}. The lattice parameters of the AlFe$_2$B$_2$ phase in the Al$_{1.2}$Fe$_2$B$_2$ sample are comparable to the reported results of arc-melted samples. Compared to the Al$_{1.2}$Fe$_2$B$_2$ sample, the lattice parameters ‘\textit{a}’ and ‘\textit{b}’ of the Al$_{1.2}$Fe$_{1.9}$Sb$_{0.1}$B$_2$ doped sample are marginally lesser, and ‘\textit{c}’ is higher. Consequently, the unit cell volume is 0.21\% less than that of the Al$_{1.2}$Fe$_2$B$_2$ sample. In the Al$_{1.2}$Fe$_{1.9}$Ge$_{0.1}$B$_2$, all three lattice parameters ‘\textit{a}’, ‘\textit{b}’ and ‘\textit{c}’ are higher, resulting in a 1.4\% increase in the lattice volume compared to un-doped Al$_{1.2}$Fe$_2$B$_2$. The Al$_{1.2}$Fe$_{1.9}$Ga$_{0.1}$B$_2$ also showed a similar change in the lattice parameter and unit cell volume.\\
\begin{table*}
\caption{Lattice parameter and secondary phases observed in Al$_{1+x}$Fe$_2$B$_2$ and Al$_{1.2}$Fe$_{0.9}$M$_{0.1}$B$_2$.}
\centering
\begin{tabular}{c c c c c c c c c}
\hline
sample & Phase & &Lattice parameters& & Imp. & Fe-Fe (\textit{ab}) & Fe-Fe (\textit{ac}) & Volume\\
 & & \textit{a} ($\AA$) & \textit{b} ($\AA$) & \textit{c} ($\AA$) & Phase & ($\AA$) & ($\AA$) & ($\AA^3$)\\
\hline
\hline
Al$_{1.2}$Fe$_2$B$_2$ & AlFe$_2$B$_2$ & 2.924(5) & 11.033(9) & 2.871(1) & FeB & 2.732 & 2.925 & 92.61\\
\hline
AlSb$_{0.1}$Fe$_{1.9}$B$_2$ & AlFe$_2$B$_2$ & 2.915(4) & 11.025(8) & 2.873(7) & FeB & 2.736 & 2.914 & 92.37\\
& & & & & Al$_{13}$Fe$_4$ & & &\\
\hline
AlGa$_{0.1}$Fe$_{1.9}$B$_2$ & AlFe$_2$B$_2$ & 2.875(4) & 11.055(9) & 2.880(8) & FeB & 2.704 & 2.928 & 91.58\\
& & & & & Al$_{13}$Fe$_4$ & & &\\
\hline
AlGe$_{0.1}$Fe$_{1.9}$B$_2$ & AlFe$_2$B$_2$ & 2.933(7) & 11.054(2) & 2.877(8) & AlB$_2$ & 2.717 & 2.937 & 93.33\\
 \hline
\end{tabular}
\label{Table I}
\end{table*}

Fig.~\ref{fig2a} shows the SEM micrograph of Al$_{1.2}$Fe$_2$B$_2$ and Al$_{1.2}$Fe$_{1.9}$M$_{0.1}$B$_2$ (M= Sb, Ge, and Ga) samples taken in the secondary electron mode over the metallographically polished surface. The Al$_{1.2}$Fe$_2$B$_2$ shows a matrix phase embedded with uniformly distributed secondary phase/s. The EDS analysis of the matrix phase shows Al, Fe and B, indicating the AlFe$_2$B$_2$ phase. The secondary phase composition analyzed showed peaks (inset Fig.~\ref{fig1a}a) corresponding to only Fe and B, indicating the FeB phase. In the Al$_{1.2}$Fe$_{1.9}$Sb$_{0.1}$B$_2$, the matrix phase shows Al, Fe and B with a small amount of Sb (up to 0.13 at.\%). The Sb is predominantly present in the secondary phase, which EDS analysis shows contains Al and Sb, with little Fe (inset Fig.~\ref{fig2a}b). The microstructure also shows particles containing only Fe and B, indicating the FeB phase (Fig.~\ref{fig2a}b).  This shows that Al$_{1.2}$Fe$_{1.9}$Sb$_{0.1}$B$_2$ has multiple secondary phases, and the Sb doping predominantly reacts with excess Al to form the AlSb impurity phase. In the sample doped with Ge, the microstructure (Fig.~\ref{fig2a}c) also consists of multiple secondary phases in the AlFe$_2$B$_2$ matrix. The matrix is relatively inhomogeneous on the microscale compared to other doped samples. Secondary phase/s of Al$_x$By composition with the high atomic percentage of B (arrow marked in Fig.~\ref{fig2a}c) have been observed throughout the microstructure. The Ge in the matrix is below the detectable limit of EDS. However, some white regions containing high Ge with Al, Fe, and B, mostly surrounding the AlxBy phase, were observed (inset Fig.~\ref{fig2a}c). The Ga-doped samples (figure 2d) showed the AlFe$_2$B$_2$ matrix phase with two types of grains (arrow marked in Fig.~\ref{fig2a}d) predominantly consisting of Al, Fe, and B with different Fe: Al ratios and the FeB secondary phase. Few particles containing Al and Fe with high percentages of Ga were also observed (inset Fig.~\ref{fig2a}d).\\ 
\begin{figure}
\centering
  \includegraphics[width=0.99\linewidth]{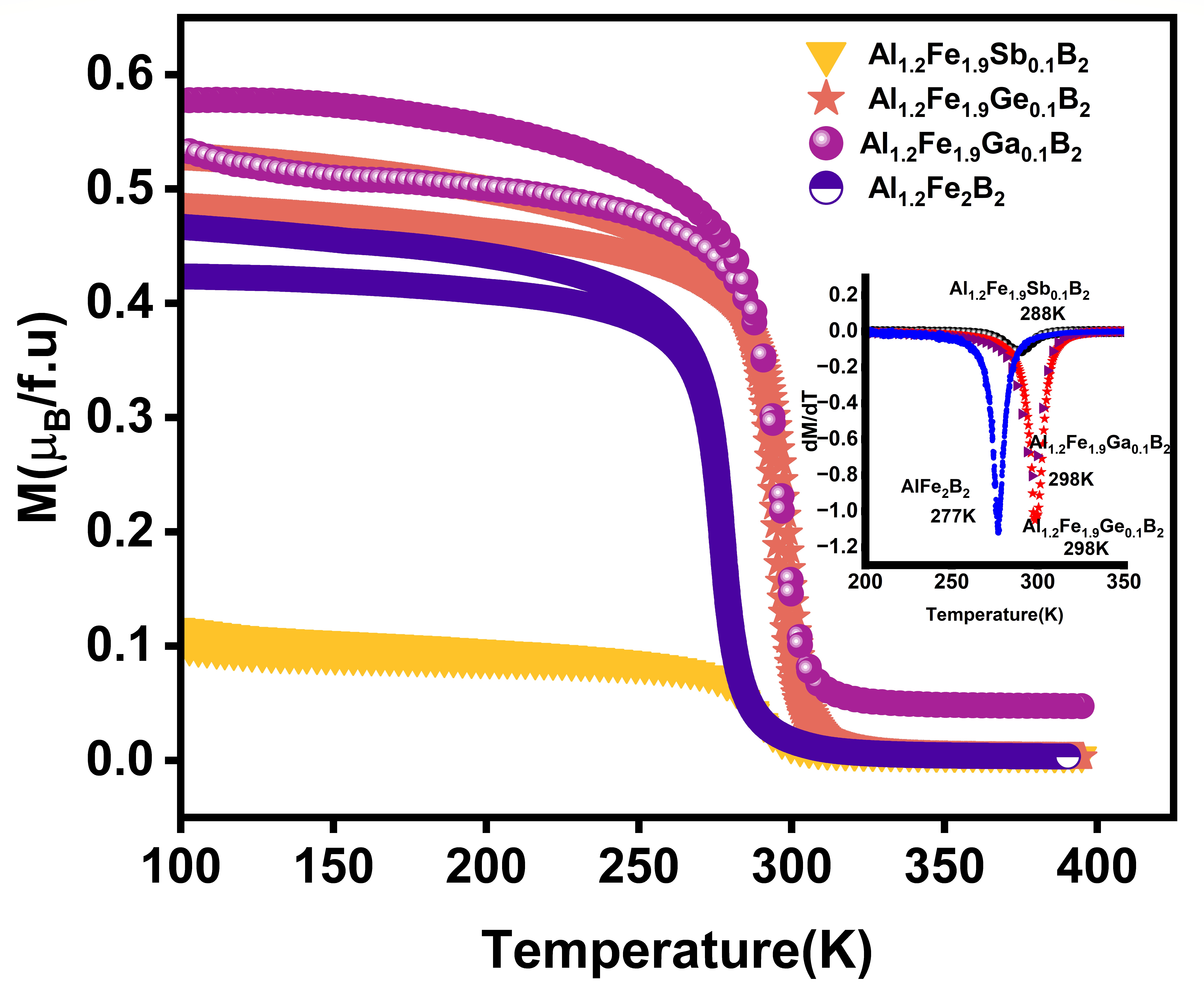}
  \caption{M vs T curves of Al$_{1.2}$Fe$_2$B$_2$ and Al$_{1.2}$Fe$_{1.9}$M$_{0.1}$B$_2$ compounds under 500 Oe with the corresponding dM/dT vs temperature in the inset.}
  \label{fig3}
\end{figure}
\begin{figure*}
\centering
  \includegraphics[width=0.9\linewidth]{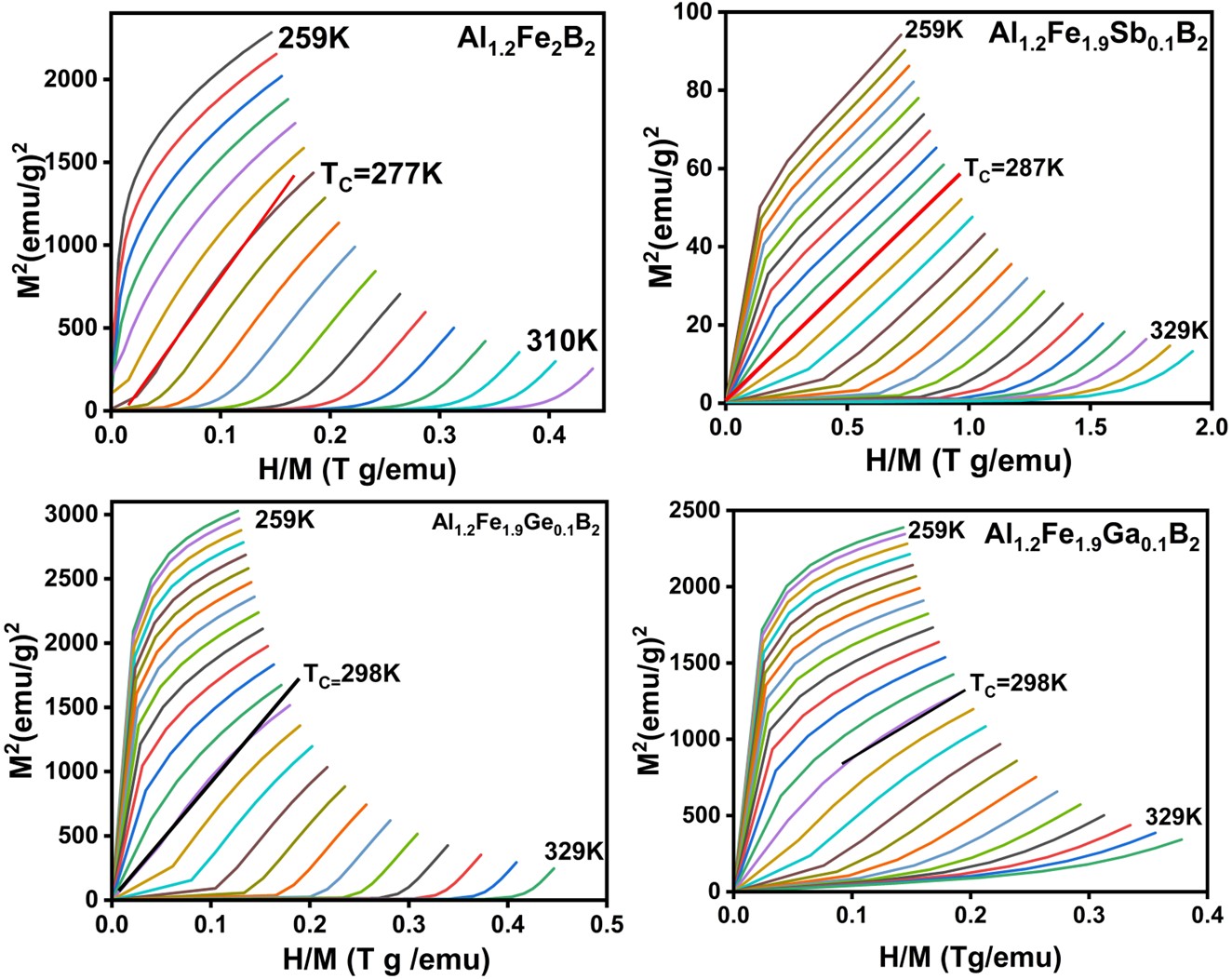}
  \caption{Arrot plots of Al$_{1.2}$Fe$_2$B$_2$ and Al$_{1.2}$Fe$_{1.9}$M$_{0.1}$B$_2$ compounds.}
  \label{fig4}
\end{figure*}
\subsection{Magnetic Properties}
Fig.~\ref{fig3} shows the magnetization-temperature (M vs. T) curves of Al$_{1.2}$Fe$_2$B$_2$ and Al$_{1.2}$Fe$_{1.9}$M$_{0.1}$B$_2$ (M=Sb, Ge and Ga) samples that were recorded between 100K and 350K under a magnetic field of 500 Oe under zero field cooling (ZFC) - field cooling (FC) mode. As shown in the figure, there are bifurcations between the ZFC and FC curves, indicating the onset of magnetic irreversibility in the ferromagnetic region. AlFe$_2$B$_2$ has a magnetic transition from the second-order ferromagnetic to the paramagnetic transition phase around room temperature. \\
The Curie temperature ($T_C$) was determined using the dM / dT plot in the inset of Figure 3. The Curie temperature of Al$_{1.2}$Fe$_2$B$_2$ is 277\,K, comparable to the 272-303\,K reported in the literature for samples prepared by the melt route.\cite{R9,R13,R15,R19,R20} With doping, the $T_C$ shifts to the higher side, and the Al$_{1.2}$Fe$_{1.9}$Ge$_{0.1}$B$_2$ and Al$_{1.2}$Fe$_{1.9}$Ga$_{0.1}$B$_2$ exhibit a significant shift, approaching 297\,K and 296\,K, respectively.  Ge doping increases $T_C$ because of its similar atomic radius and electronic structure, which enhances the ferromagnetic exchange interactions. Ga doping also raises $T_C$ by stabilizing the crystal structure and reducing magnetic anisotropy. Sb doping exhibits only moderate improvement of 10\,K in $T_C$.  
An interesting aspect of the observed doping effect is that in Al$_{1.2}$Fe$_{1.9}$Sb$_{0.1}$B$_2$, the M vs. T differ substantially from those of other compounds. The magnetization of Al$_{1.2}$Fe$_2$B$_2$ increases with Ge and Ga doping. In contrast, the Sb-doped sample exhibits a sharp decrease in magnetization. A study by Du et al. reported similar behavior in Mn-substituted AlFe$_2$B$_2$ lattice parameters, and the magnetization decreases substantially with Mn percentage.\cite{R21} By comparing the structural parameters, it was found that the lattice parameters\textit{ a} and \textit{b} of Al$_{1.2}$Fe$_{1.9}$Sb$_{0.1}$B$_2$ decrease, whereas the \textit{c}, the separation between Fe-Fe along the c-axis is marginally higher than the parent compound. In the orthorhombic structured AlFe$_2$B$_2$, the magnetic properties are highly sensitive to the Fe-Fe distance in the \textit{a-b} plane. Decreasing the distance between Fe-Fe in the ‘\textit{a}’ and ‘\textit{b}’ directions can dramatically decrease the magnetic interaction between the Fe atoms below or above 2.8$\AA$, where the exchange interaction is maximum. According to Heisenberg exchange interactions, decreasing ‘\textit{a}’ and ‘\textit{b}’ favors antiferromagnetic interaction, whereas enhancement in \textit{a} and \textit{b} favors ferromagnetic interaction. The lattice parameters of Al$_{1.2}$Fe$_{1.9}$Sb$_{0.1}$B$_2$, a (representing the distance between Fe-Fe along the a-axis), and b are decreasing with respect to parent Al$_{1.2}$Fe$_2$B$_2$, indicating Sb doping favors antiferromagnetic interaction between the Fe atoms. However, in Al$_{1.2}$Fe$_{1.9}$Ge$_{0.1}$B$_2$ and Al$_{1.2}$Fe$_{1.9}$Ga$_{0.1}$B$_2$, lattice parameters \textit{a} and \textit{b} are higher than Al$_{1.2}$Fe$_2$B$_2$ (Table~\ref{Table I}), resulting in higher magnetization. It is to be noted that, unlike Al$_{1.2}$Fe$_{1.9}$Ga$_{0.1}$B$_2$, which shows non-zero magnetization due to FeB ferromagnetic impurity, Sb doping has relatively less FeB phase and the M (T) is close to zero above $T_C$. The AlSb present is a ferromagnetic phase; hence, its effect on magnetization is likely to be insignificant.\cite{R22} Hence, the decrease in M in the Sb-doped sample is primarily attributed to the antiferromagnetic interaction between Fe-Fe. \\ 
Fig.~\ref{fig4} shows the Arrott plot of AlFe$_2$B$_2$ and AlFe$_{1.9}$M$_{0.1}$B$_2$. The positive slope of the Arott plot indicates that the material undergoes SOMTM, which is more suitable for magnetocaloric applications.\\
\begin{figure*}
\centering
  \includegraphics[width=0.9\linewidth]{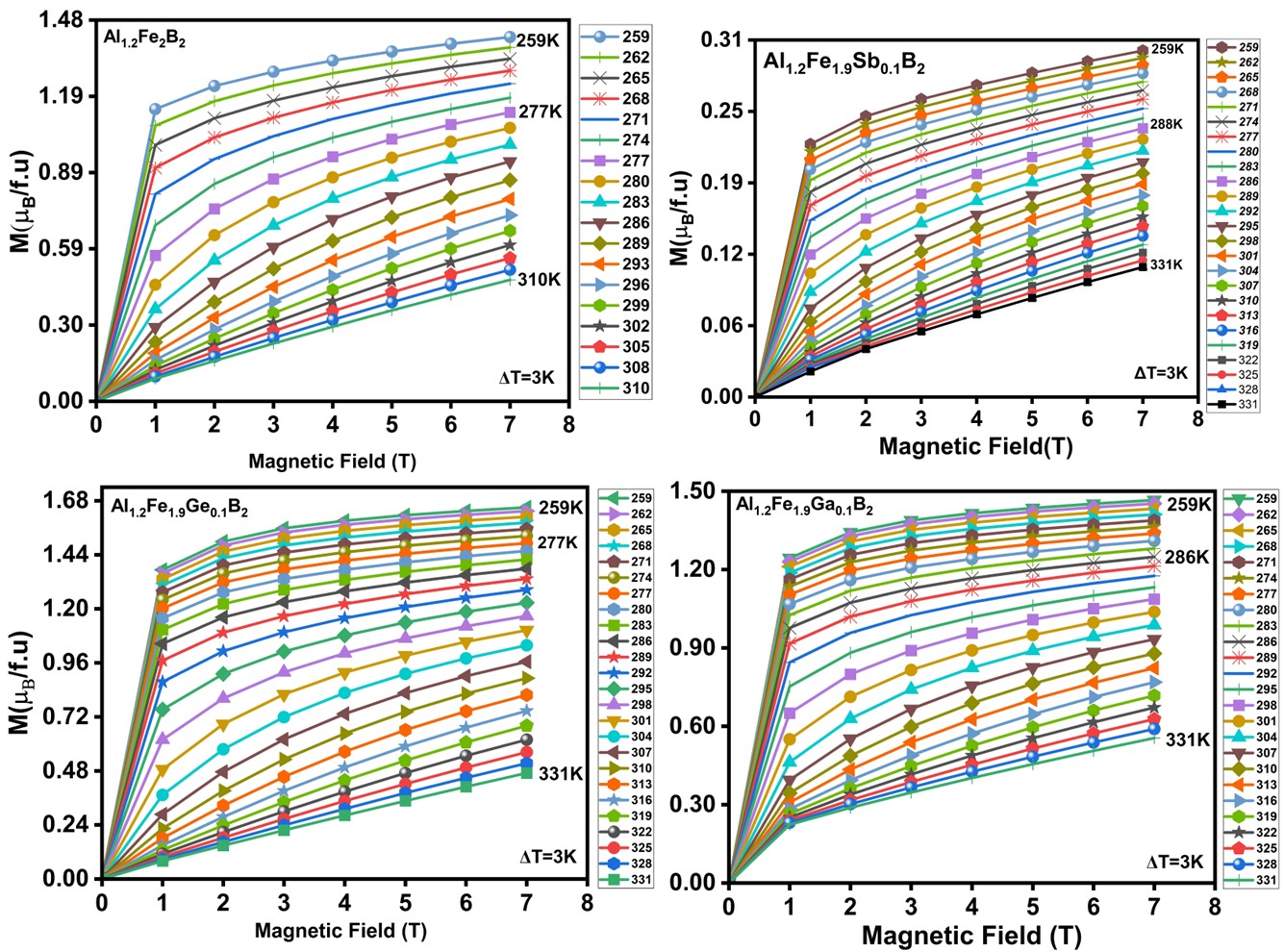}
  \caption{Isothermal magnetization curve of Al$_{1.2}$Fe$_2$B$_2$ and Al$_{1.2}$Fe$_{1.9}$M$_{0.1}$B$_2$ (M=Ga, Sb, Ge)}
  \label{fig5}
\end{figure*}
\begin{table*}
\caption{Maximum $\Delta S_m$, $T_{FWHM}$, and the RCP obtained for Al$_{1.2}$Fe$_2$B$_2$ and Al$_{1.2}$Fe$_{1.9}$M$_{0.1}$B$_2$ compounds. $*$- as cast, $**$- pure phase (acid treated to remove Al$_{13}$Fe$_4$)}
\centering
\begin{tabular}{c c c c c c c}
\hline
sample & -$\Delta S_m $& & $T_{FWHM}$& & RCP (J/kg) & \\
       & 2 T & 5 T& 2T & 5T & 2T & 5T\\
\hline
\hline
Al$_{1.2}$Fe$_2$B$_2$ & 2.31& 4.28 & 37.26 & 52.52 & 85.92 & 224.89\\
\hline
Du et. al.\cite{R10} & 3.4& 7.2 & - & - & 74 & 216\\
\hline
Tan et al\cite{R8}& 4.4& 7.3 &-  & - & 88 & 210\\
\hline
Barua et al. \cite{R12}& 2.7& -&- & -&- & -\\
\hline
Lee at al \cite{R26}$^*$& 1.85& 4.0 & 31.59 & 39.53 & 58.53 & 157.97\\
\hline
Lee et al. \cite{R26}$^{**}$& 3.07 & 6.49 & 23.99 & 29.61 & 73.77 & 192.17\\
\hline
Limichhane et al. \cite{R27}& 3.78 &- & -& - & - & -\\
\hline
 Al$_{1.2}$Fe$_{1.9}$Ge$_{0.1}$B$_2$ & 2.55& 4.75 & 29.56 & 44.37 & 75.47 & 210.62\\
\hline
 Al$_{1.2}$Fe$_{1.9}$Ga$_{0.1}$B$_2$  & 1.92& 3.55 & 30.94 & 46.37 & 59.45 & 164.96\\
\hline
 Al$_{1.2}$Fe$_{1.9}$Sb$_{0.1}$B$_2$  & 0.31& 0.61 & 34.76 & 77.75 & 10.91 & 47.19\\
\hline
\end{tabular}
\label{Table II}
\end{table*}

\begin{figure}
\centering
  \includegraphics[width=0.99\linewidth]{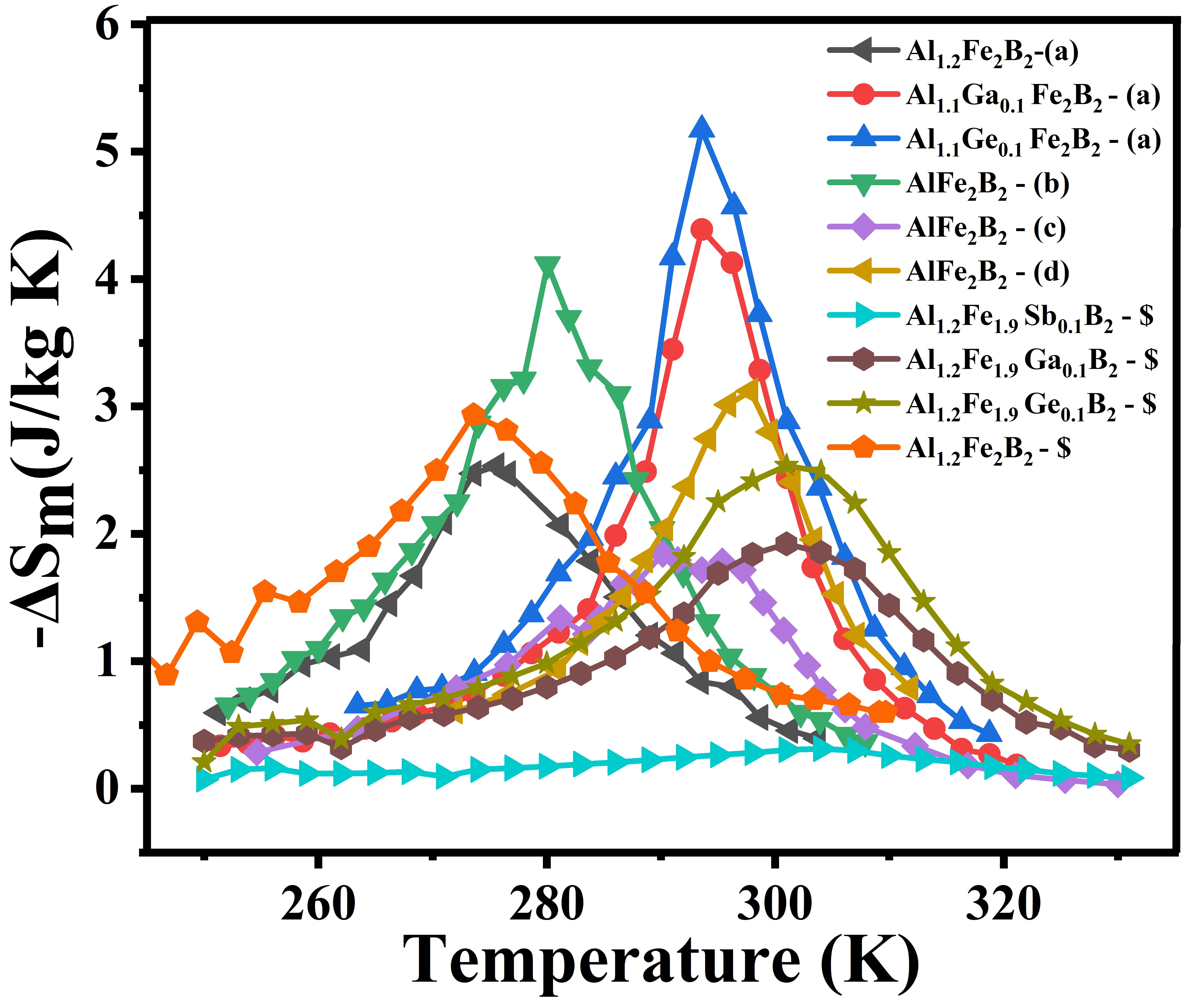}
  \caption{Magnetic entropy change with temperature in Al$_{1.2}$Fe$_2$B$_2$ and Al$_{1.2}$Fe$_{1.9}$M$_{0.1}$B$_2$ and data replotted from other reported work for similar compositions, (a)-\cite{R12}, (b)-\cite{R8}, (c)-\cite{R25}, (d)-\cite{R17}, \$-Present work}
  \label{fig6}
\end{figure}

Fig.~\ref{fig5} shows M$^2$ vs H/M isotherms measured at different temperatures in the step size of 3\,K from 1-7 T below $T_C$. The M vs. T curve of Al$_{1.2}$Fe$_2$B$_2$ exhibits ferromagnetic behavior and reaches close to saturation at the maximum field of 7 T used. It is observed that with Ge and Ga doping, the saturation in magnetization increases for an applied field. However, with Sb doping, the magnetization is far from saturation, even at a 7 T field. With temperature beyond $T_C$, the saturation gradually decreases and becomes almost linear, a characteristic of paramagnetic behavior.  
The magnetic entropy change (-$\Delta$S$_m$) associated with ferromagnetic to paramagnetic transformation in these materials was calculated from magnetization isotherms recorded in the range of 240-330 K. Fig.~\ref{fig6} presents the magnetic entropy change with temperature (-$\Delta$S$_m$ vs. T) under 2T field obtained in Al$_{1.2}$Fe$_2$B$_2$ and Al$_{1.2}$Fe$_{1.9}$M$_{0.1}$B$_2$ and reported data for these compounds under same applied field. The -$\Delta$S$_m$ values are calculated by Maxwell’s relation, which follows as:
\begin{equation}
    \Delta S_m(T)=\int\frac{dM}{dT}HdH
\end{equation}
The peak -$\Delta$S$_m$, temperature range at full-width half maximum in $\Delta$S$_m$ vs.T curve ($T_{FWHM}$) and relative cooling power (RCP) of materials tested in this work and other reported values are listed in Table~\ref{Table II}. The Al$_{1.2}$Fe$_2$B$_2$ showed a maximum $\Delta$S$_m$ of 2.3 J/kg. K, comparable to the 2.7 J/kg. K reported a similar composition.\cite{R12} With doping, it is found that the -$\Delta$S$_m$ peak increases marginally with Ge; however, with Ga and Sb-doping, the magnetic entropy decreases substantially. In a work by Barua et al. in melt and annealed samples, it was observed that with 0.1\% Ga and Ge doping -$\Delta$S$_m$ values are increasing in Al$_{1.1}$Fe$_2$B$_2$.\cite{R12,R13,R14,R15,R16,R17,R18,R19,R20,R21,R22} In the present work with 20\% excess Al, the -$\Delta$S$_m$ value of Ga-doped samples decreases, and Ge exhibits marginal improvement. Both Ge and Ga have very low solubility in the AlFe$_2$B$_2$ matrix. The increase in $\Delta$S$_m$ with these dopants was attributed to Fe anti-site defect where the Fe atoms occupy the Al position. The absence of notable improvement in $\Delta$S$_m$ with Ga and Ge doping is possibly due to the absence of these Fe antisite defects. The stoichiometry Al$_{1.2}$Fe$_{1.9}$Ge$_{0.1}$B$_2$ and Al$_{1.2}$Fe$_{1.9}$Ga$_{0.1}$B$_2$ chosen in this work, in which the Fe is sub-stoichiometric compared to ideal (Al: Fe: B of 1:2:2), appears to have less Fe in Al sites. From the findings of Lejeune et al., it appears that the $\Delta$S$_m$ peak and FWHM were noticeably influenced by the Fe: Al ratio.\cite{R23} In the doped samples, though, the AlFe$_2$B$_2$ stable phase forms are deficient in Fe. The secondary phase fraction and microstructural inhomogeneity in the doped samples, particularly in Ga, can also be contributed to low entropy value. The Sb-doping dramatically reduces the AlFe$_2$B$_2$ magnetization (only 13.4\% of the parent compound), which is reflected in the peak value of $\Delta$S$_m$.\\  
\begin{figure}
\centering
  \includegraphics[width=0.99\linewidth]{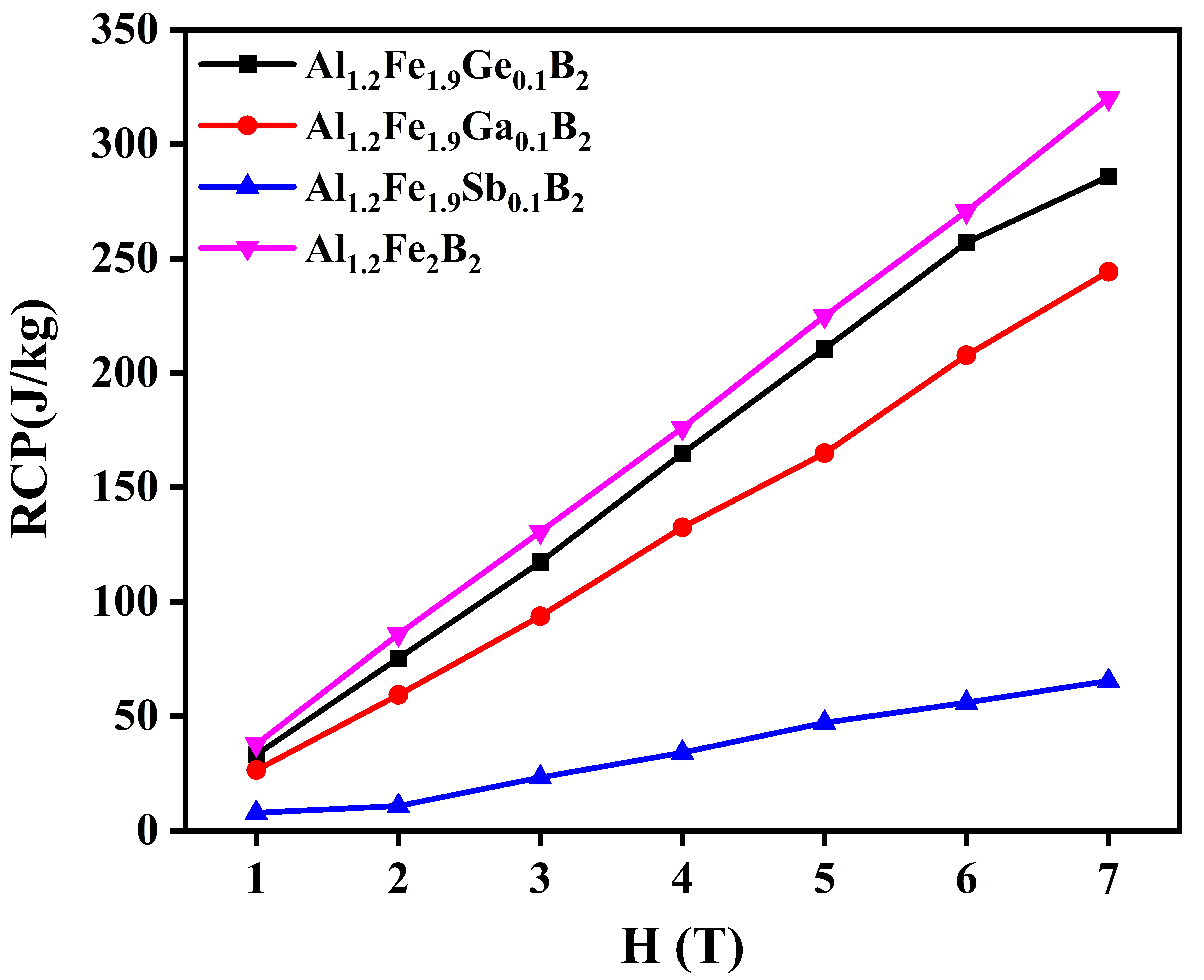}
  \caption{RCP of the Al$_{1.2}$Fe$_2$B$_2$ and Al$_{1.2}$Fe$_{1.9}$M$_{0.1}$B$_2$ estimated under the varying applied external field}
  \label{fig7}
\end{figure}
The refrigeration capacity of the magneto-caloric materials is estimated by relative cooling power (RCP). It is a key parameter used to quantify the effectiveness of the MC material's ability to produce a temperature change when subjected to a changing magnetic field. RCP is calculated by,
\begin{equation}
    RCP=[-\Delta S_m] \times T_{FWHM}
\end{equation}
In the present work, RCP is calculated from 1-7 T, and the values obtained are given in Fig.~\ref{fig7}. The RCP of Al$_{1.2}$Fe$_2$B$_2$ under 2 T is comparable to other reported values; with a higher applied field of 5T, it was the highest among the results reported so far [16, 24]. In the Ge-doped sample, the RCP is marginally lesser than the parent compound. However, in Ga and Sb doping, the RCP was reduced significantly mainly due to lower -$\Delta$S$_m$. The Ge and Ga doped samples showed $T_{FWHM}$ narrower than the parent compound, but the -$\Delta$S$_m$ peaks at 19-20\,K above the peak of AlFe$_2$B$_2$. This shows that by combining these materials, stable refrigeration over a broad temperature range is possible. The Sb-doped sample showed a very low RCP over the entire 1-7\,T applied field, which is unsuitable for MC application.\\
\section{Conclusion}
Al$_{1.2}$Fe$_2$B$_2$ doped with Sb, Ge and Ga prepared by vacuum arc melting was investigated for structural, magnetic and magnetocaloric properties. The lattice parameter of the Sb doped sample changes anisotropically with \textit{a}, \textit{b} decrease, and \textit{c} increase, whereas in Ge and Ga, all the parameters increase. Doping generates additional secondary phases apart from FeB, such as AlSb in Al$_{1.2}$Fe$_{1.9}$Sb$_{0.1}$B$_2$ and AlB$_2$ in Al$_{1.2}$Fe$_{1.9}$Ge$_{0.1}$B$_2$. The Curie temperature of 277 K of Al$_{1.2}$Fe$_2$B$_2$ with Ge and Ga doping increased to 297\,K and 296\,K, respectively. The Sb showed a moderate improvement in $T_C$ to 287 K. The relative cooling power (RCP) of the doped materials, particularly Ga and Sb doped, is significantly less than that of the parent compound. The low relative cooling power of the doped compounds can be attributed to factors like additional impurity phases apart from FeB and sub-stoichiometry of Fe from ideal (Al: Fe of 1:2), which affects the magnetic entropy change to a large extent.\\
\section{Conflicts of interest}
There are no conflicts to declare.
\section*{Acknowledgements}
The authors would like to acknowledge the funding of the Department of Science and Technology, Government of INDIA, for the Technical Research Centre Grand. The authors are grateful to the Director, ARC-I, for his support.\\
%
%


%
\end{document}